\documentclass[12pt]{article}

\def\beq{\begin{eqnarray}}
\def\eeq{\end{eqnarray}}
\def\bea{\begin{eqnarray}}
\def\eea{\end{eqnarray}}

\def\tev{\, {\rm TeV}}

\newcommand{\gsim}{\lower.7ex\hbox{$\;\stackrel{\textstyle>}{\sim}\;$}}
\newcommand{\lsim}{\lower.7ex\hbox{$\;\stackrel{\textstyle<}{\sim}\;$}}

\newcommand{\drawsquare}[2]{\hbox{%
\rule{#2pt}{#1pt}\hskip-#2pt
\rule{#1pt}{#2pt}\hskip-#1pt
\rule[#1pt]{#1pt}{#2pt}}\rule[#1pt]{#2pt}{#2pt}\hskip-#2pt
\rule{#2pt}{#1pt}}

\newcommand{\fund}{\raisebox{-.5pt}{\drawsquare{6.5}{0.4}}}
\newcommand{\antifund}{\overline{\fund}}

\begin{document}

\setlength{\baselineskip}{0.25in}

\begin{titlepage}
\noindent
\begin{flushright}
MIFP-06-22 \\
\end{flushright}
\vspace{1cm}

\begin{center}
  \begin{Large}
    \begin{bf}
Hidden Sector Baryogenesis \\

    \end{bf}
  \end{Large}
\end{center}
\vspace{0.2cm}
\begin{center}
\begin{large}
Bhaskar Dutta and Jason Kumar \\
\end{large}
  \vspace{0.3cm}
  \begin{it}
Department of Physics \\
Texas A\&M University \\
        ~~College Station, TX  77840, USA \\
\vspace{0.1cm}
\end{it}

\end{center}

\begin{abstract}

We introduce a novel mechanism for baryogenesis, in
which mixed anomalies between the hidden sector
and $U(1)_{baryon}$ drive the baryon asymmetry.  We
demonstrate that this mechanism occurs quite naturally
in intersecting-brane constructions of the Standard
Model, and show that it solves some of the theoretical
difficulties faced in matching baryogenesis to experimental
bounds. We illustrate with a specific example model. We 
also discuss the possible signals at the LHC.

\end{abstract}

\vspace{1cm}

\begin{flushleft}
August 2006
\end{flushleft}

\end{titlepage}

\setcounter{footnote}{0}
\setcounter{page}{2}
\setcounter{figure}{0}
\setcounter{table}{0}


\section{Introduction}

One of the great puzzles facing theoretical
physics is the question of why we observe so
little anti-matter, as compared to matter~\cite{Steigman:1976ev}.
The generation of the asymmetry is called
baryogenesis, and Sakharov
demonstrated~\cite{Sakharov:1967dj} that
the three conditions required for it to occur
are
\begin{itemize}
\item{Violation of baryon number, $B$}
\item{$C$ and $CP$ violation}
\item{Departure from thermal equilibrium}
\end{itemize}
Several ideas have been proposed to satisfy these
conditions, the most prominent of which are
GUT baryogenesis,
Affleck-Dine baryogenesis~\cite{Affleck:1984fy},
baryogenesis via leptogenesis~\cite{Fukugita:1986hr}
and electroweak baryogenesis~\cite{EWBG}.  However all of
these models have various theoretical or
experimental constraints which make the fit
to data problematic.

A parallel thread in model-building has been the
construction, within the context of string theory,
of intersecting brane models
(IBMs)~\cite{IBM,confine,MS}
which can perhaps provide a description of real-world low
energy physics.  These stringy constructions have
provided new insights into the types of
beyond-the-Standard-Model physics one might expect to find
at colliders.

One of the ubiquitous features of these IBMs is
the existence of hidden sectors, arising from the
gauge theory living on extra branes (above and
beyond the visible
sector branes needed to generate the Standard Model).
It is rather generic in this context for the
hidden sector groups to have mixed anomalies with
Standard Model $U(1)$'s.
We suggest here a mechanism in which mixed
anomalies between baryon number, $U(1)_B$, and
hidden sector gauge groups can drive
baryogenesis.

This provides a new mechanism for baryogenesis which
not only provides a unique phenomenological signature,
but also seems to appear rather generically in a large
class of stringy constructions.

\section{Motivation}

In intersecting brane models, one
obtains a Standard Model gauge theory as the low-energy
limit of the theory of open strings which begin and end
on a set of intersecting D-branes (the ``visible sector").
In such models,
however, one generically can find additional D-branes,
and the gauge theory living on those branes provides a
hidden sector.
The strings which stretch between those
hidden sector branes, and between hidden and visible sector
branes, form exotic matter\footnote{Since there can be exotic
matter charged under both the hidden and Standard Model gauge
groups, our hidden sector is more precisely a pseudo-hidden
sector.}.

In a construction with $N=1$ supersymmetry (which arises
in orientifold models), the strings stretching between branes
yield degrees of freedom which are arranged into chiral multiplets.
The net number of bifundamental chiral
multiplets\footnote{Other
representations arise when the effects of the orientifold are
accounted for, but we will not need them here.}
stretched between two
branes~\cite{Berkooz:1996km}
is given by the topological intersection number $I_{ab}$:
\bea
I_{ab}\, \, {\rm multiplets} \rightarrow (\fund_a, \antifund_b)
\eea
Nontrivial topological intersections are somewhat generic in many
constructions.  One of the best understood IBMs arises from
toroidal orientifolds (for example, $T^6/Z_2 \times Z_2
(\times \Omega R)$ ) of Type IIA string theory, on which
D6-branes are wrapped.  The branes in this example wrap
3-cycles on the compact dimensions, which can be represented
by three coprime ordered-pairs of wrapping numbers:
$(n_1 ,m_1)(n_2 ,m_2)(n_3 ,m_3)$, where $(n_i ,m_i)$ are the
wrapping numbers on the $a$ and $b$ cycles of the $i$th torus.
The topological intersection between branes $a$ and $b$ is
then
\bea
I_{ab} = \prod_{i=1} ^3 (n_i ^a m_i ^b - m_i ^a n_i ^b)
\eea
It is clear that $I_{ab}=0$ only if D6-branes $a$ and $b$
have the same wrapping numbers on at least one torus.  For a
generic choice of wrapping numbers, this will not be the case,
and $I_{ab} \neq 0$.

For a more general manifold, an orthogonal basis of 3-forms may be
written as $\alpha_i$, $\beta^j$, where $\int_{\Sigma} \alpha_i \wedge
\beta^j =\delta_i ^j$.  Without loss of generality, assume brane $a$
wraps the 3-cycle dual to $\alpha_1$.  The form dual to the cycle
wrapped by brane $b$ is
\bea
\gamma = \sum_i a^i \alpha_i + b_i \beta^i .
\eea
In this case, $I_{ab}=0$ only if $b_1 =0$.  For a generic choice of
$a_i$, $b^j$, this will not be the case, and again $I_{ab} \neq 0$.

It is necessary to ensure that the gauge theory has canceled
anomalies.  The cancelation of cubic anomalies is automatically
ensured by the RR-tadpole constraints (i.e, the constraint that
all space-filling charges cancel).  There can also be mixed anomalies,
however, which are canceled by a generalized Green-Schwarz mechanism.
If a symmetry is broken by such an anomaly, then the associated gauge
boson will receive mass through the Steuckelberg mechanism, and the
symmetry will appear to be an anomalous global symmetry at low-energies.
If two branes $a$ and $b$
have non-trivial intersection, then there will be chiral fermions
transforming under the groups $G_a$ and $G_b$, where $G_{a,b}$
are the gauge groups living on branes $a$ and $b$ respectively.
It is clear from the field theory analysis that there will thus
be a $U(1)_a - G_b ^2$ mixed anomaly (where $U(1)_a$ is the diagonal
$U(1)$ subgroup of $G_a$) given by
\bea
\partial_{\mu} j_a ^{\mu} &=& {I_{ab}\over 32\pi^2}
Tr F_b \wedge F_b
\eea

In a large class of intersecting brane world models, $SU(3)_{qcd}$ arises
as a subgroup of a $U(3)$ gauge group living on a stack of 3 parallel
D-branes (in certain cases where there is an orientifold plane, there will
actually be 6 parallel D-branes in this stack).
In such cases, the charge under the diagonal $U(1)_B$ is baryon number.  As
we have seen, $U(1)_B$
will generically have mixed anomalies with other gauge groups (both visible
and hidden sector), provided that the $U(3)_{qcd}$ stack of branes and the
other stack have non-trivial intersection.

We will consider the case where there is an anomaly between the hidden sector
and $U(1)_B$.  As a result, the divergence of the baryon current will be given
by
\bea
\partial_{\mu} j_B ^{\mu} \sim Tr F \wedge F
\eea
where $F$ is the field strength of the hidden sector gauge theory.
Instantons or sphalerons in the hidden sector will then violate
baryon number, providing a source for the baryogenesis.

\section{Baryogenesis driven by the hidden sector}

Having motivated this mechanism from intersecting brane world
constructions, we will develop this idea from the point
of view of the low-energy effective field theory.  In fact, motivation
aside, this mechanism can appear just as readily in non-stringy constructions,
and it will be easier to find specific models in low-energy effective
field theory. We refer to Ref.\cite{related} for other work on
baryogenesis in related contexts.

We will consider a theory with $N=1$ SUSY and Standard Model gauge group
and matter content, as well as a non-trivial hidden sector including
hidden group $G$.
We will need four features:
\begin{itemize}
\item{The cancelation of all cubic anomalies (in IBMs, this is ensured
by the RR-tadpole constraints}
\item{A non-vanishing $U(1)_B - G^2$ mixed anomaly}
\item{Vanishing $U(1)_Y$ mixed anomalies}
\item{A Yukawa coupling which permits exotic baryons to
decay to SM baryons}
\end{itemize}
All multiplets charged under the fundamental of $SU(3)_{qcd}$ have
charge ${1\over 3}$ under $U(1)_B$.  In an intersecting brane model
this will arise naturally, as $U(1)_B = {1\over 3}U(1)_{diag}$ is a
gauged subgroup of $U(3)_{qcd}$.  In a more general field theory model,
$U(1)_B$ arises simply as a global symmetry.

The vanishing of $U(1)_Y$ mixed anomalies is easy to arrange in
intersecting brane models.  In that case, $U(1)_Y$ arises as a linear
combination of $U(1)$'s, and in many constructions it is easy to arrange
for the existence of such a non-anomalous symmetry.
In such constructions, the vanishing of the hypercharge anomaly naturally
leads to the existence of the appropriate
Yukawa coupling.  For example, one might arrange for the $U(1)_Y - G^2$
anomaly to vanish by ensuring a non-trivial intersection between the $G$
branes and a $U(1)_{T_{3R}}$ brane.  But as we will see, this 
permits a Yukawa coupling
which allows exotic quarks to decay to right-handed quarks, plus an exotic
scalar.
From the effective
field theory point of view, we merely need to choose our exotic with matter
with appropriate hypercharge couplings to ensure vanishing anomalies and
the appropriate Yukawa couplings.

\subsection{A specific model}

We will now look at a specific model with a hidden gauge group
$G$ contained in a larger hidden sector.
In our model, we have $U(1)_Y = {1\over 2}(U(1)_B -U(1)_L
+U(1)_{T_{3R}} -U(1)_G +...)$, where $U(1)_G$ is the diagonal $U(1)$ subgroup
of $G$.  We have 2 chiral
multiplets $q_i$ transforming in
the bifundamental of $(U(3)_B,G)$ and with hypercharge $Q_Y = {2\over 3}$;
four multiplets $\lambda_j$ transforming in the fundamental of
$G$ with charge $Q_{T_{3R}}=-1$ and hypercharge $Q_Y =-1$; one chiral multiplet
$\eta$ transforming in the fundamental of $G$ with charge
$Q_{T_{3R}}=1$ and hypercharge $Q_Y =0$; and one chiral multiplet $\xi$
transforming in the anti-fundamental of $G$ with charge
$Q_L =1$ and hypercharge
$Q_Y =0$. The charges for this specific model are described in Table 1.  This could arise in a brane model (assuming we label the
branes as follows: $a=U(3)_B$, $b=U(1)_{T_{3R}}$, $c=U(1)_L$ and $g=G$) with
intersection numbers $I_{ag}=2$, $I_{gb}=4$, $I_{gb'}=1$, and
$I_{cg}=1$, where $b'$ is the orientifold image of the $b$ brane.

\begin{table}
\begin{center}
\caption{Particle spectrum for the example model.}
\begin{tabular}{|r|r|r|r|r|r|}
  \hline
  particle & $Q_B$ & $Q_G$ & $Q_{T_{3R}}$ & $Q_L$ & $Q_Y$ \\
  \hline
  $q_i$ & ${1\over 3}$ & -1 & 0 & 0 & ${2\over 3}$\\
  $\lambda_j$ & 0 & 1 & -1 &0 &-1\\
  $\eta$ & 0 & 1 & 1 & 0 &0\\
  $\xi$ & 0 & -1 & 0 & 1 &0\\
  \hline
\end{tabular}
\end{center}
\end{table}

We see that all of this matter is charged under the
(anti)fundamental of $G$, and the net hypercharge of this matter
content is zero.  As a result, we induce no $U(1)_Y -G^2$ mixed anomaly.
Note however, that $U(1)_B$ and $U(1)_{B-L}$ have mixed anomalies with
$G$.  We assume that the rest of the hidden sector cancels the RR-tadpoles.
This ensures the cancelation of all cubic anomalies.  Assuming that there
are no symmetric or antisymmetric representations of $SU(3)_{qcd}$ (which
is easy to arrange by a judicious choice of the QCD branes), this also
ensures that there are no net chiral exotics.

The divergence of the baryon current will contain
a hidden-sector contribution given by
\bea
\partial_{\mu} j_B ^{\mu} &\propto& {1\over 32\pi^2}(g_G ^2 Tr\,F_G
\wedge F_G  +... )
\eea
At low energies
$U(1)_B$ will appear to be an anomalous global symmetry.
We will assume that $G$ and other hidden sector gauge groups break at some
scale (breaking
or confinement will be necessary in order to avoid exotic massless fermions
which are charged under the Standard Model).  The breaking of $G$ can
involve complicated hidden sector dynamics
such as brane recombination~\cite{recomb,MS}
in which other hidden sector
groups simultaneously break.  For a phenomenologically viable model,
however, the $U(1)$ hypercharge must remain
unbroken.

\subsection{Phase transitions and baryogenesis}

Having discussed the basic setup, one can now address the way
baryogenesis actually occurs.  As the universe expands and cools,
we assume that there is a phase transition (such as the spontaneous
symmetry breaking of $G$) at some temperature $T_C$.
If this $G$ phase transition is strongly first-order, then it will
result the nucleation of expanding bubbles of a broken symmetry
vacuum.

At the bubble walls, there will be a
departure from equilibrium.
During this process, $CP$ can generically be
violated in the $G$-sector.  $G$-sphalerons correspond to transitions
from one vacuum of the hidden sector theory to another, and the mixed
$U(1)_B -G^2$ anomaly implies that these transitions are accompanied by
a discrete violation of baryon number~\cite{'tHooft:1976fv}.
All of the Sakharov conditions are thus satisfied, and a baryon asymmetry can
be produced during the phase transition.
$G$-sphalerons will be unsuppressed above the phase transition, but will
generally be suppressed at low temperatures~\cite{sphaleron}.
Thus, to avoid washout of the produced
asymmetry by $G$-sphalerons after the phase transition, one must
demand the usual condition~\cite{Bochkarev:1987wf}
\bea
{v(T_c) \over T_c} \geq 1,
\eea
where $v(T_c)=\langle \phi \rangle$ is the order parameter of the first
order phase transition.
This mechanism is reminiscent of electroweak baryogenesis~\cite{Huet:1995sh},
but does not
suffer from the tunings required to fit electroweak baryogenesis into the
parameter space allowed by LEP-II data and
EDM bounds~\cite{LEPII,Barate:2003sz,EDM}.

It is interesting to note that the amount of chiral matter charged under
$G$ and $U(1)_L$ need not be the same as the amount charged under
$G$ and $U(1)_B$.  As a result, 
there may be a $U(1)_{B-L}-G^2$ anomaly (indeed, $U(1)_L$ may have no 
anomaly).
In IBMs, one expects 
a $U(1)_{B-L}-G^2$ anomaly 
unless $U(1)_L$ lives on a lepton brane which is parallel to the QCD
branes, as in a Pati-Salam model.  On thus expects that
these $G$ sphalerons can violate both
$B$ and $B-L$.  As a result, even if $G$ breaks at a scale significantly
larger than $\tev$, electroweak sphalerons will not wash out the baryon
asymmetry.  This naturally avoids one of the difficulties of GUT baryogenesis.

Of course, one can choose models where the $G$-sphaleron does preserve $B-L$.
This will occur if 
the $U(1)_L -G^2$ anomaly has the same magnitude as the 
$U(1)_B -G^2$ anomaly, 
as is the case in Pati-Salam constructions of the SM sector (where $U(1)_{B-L}$ is
a non-diagonal subgroup of $U(4)$).  In this
case, the hidden sector drives baryogenesis only if the scale of $G$ breaking
is approximately at or below the electroweak scale.  This would be natural in
a scenario where supersymmetry breaking is communicated to both the $G$ and
SM sectors by gravity/moduli.

In our specific example,
the exotic particles generated by the $G$ sphalerons will be the
exotic baryons schematically represented by $q_i q_i q_i$, as well as
$\lambda_i$, $\eta$ and $\xi$.  We will use a tilde to represent the
scalar of the appropriate chiral multiplet, while the fermion will
be represented without a tilde where no confusion is caused.
In order to provide realistic
baryogenesis, there must be a process whereby the $qqq$ baryons decay to
Standard Model baryons and the $\lambda$ fermions decay to Standard Model
particles.  Generically, there will be Yukawa coupling terms
of the form\footnote{The subscripts on $u^c$ and $e^c$ denote flavors of
right-handed up-type quarks and electron-type leptons.}
\bea
W_{yuk.} &=& c_{ik} q_i u_k ^c \eta + d_{jm} \lambda_j e_m ^c \xi +...
\eea
which allow an exotic $q_i$ quark to decay to $u^c$ and
$\tilde \eta$ and allow $\lambda_i$ to decay to
$e^c$ and $\tilde \xi$; we assume that these decays are kinematically
allowed (if this is not
the case, then we would instead find exotic baryons which do not
decay to SM baryons).
These decays conserve $R$-parity if we assign the following
charges: $Q_{\eta} = Q_{\lambda}=-1$, $Q_q=Q_{\xi}=1$.
Indeed, we must be sure that $\lambda_i$ can decay to only charged
SM and neutral exotics before
nucleosynthesis, in order to avoid $Li^6$ production
bounds~\cite{Li6}.

The fields $\eta$ and $\xi$ can play the role of dark matter particles.
These fields can get
Majorana masses once the $U(1)_{T_{3R}} \times U(1)_L$ symmetry is broken
(leaving only $U(1)_Y$).  The fermionic
parts of the fields (assuming they are lighter) can take part
in constituing the dark matter of the universe. The annihilation of
these new particles can happen via a $t$-channel exchange of exotic quarks.

But in a more general scenario where $U(1)_L$ is unbroken, $\xi$ cannot
obtain a Majorana mass.  As $\xi$ is produced by the same $G$-sphalerons, 
one expects
the $\xi$ number density to be related to the baryon number density (the precise ratio depends on
the specifics of a model).  In a
simple scenario the mass of  $\xi$  could be  $10 m_{proton}$ which  would provide a
nice mechanism for relating the baryon and dark matter densities, along the lines
of \cite{BandDM}.

As we have not specified the precise nature of the hidden sector, it is
not clear whether baryogenesis is dominated by local or nonlocal processes.
If nonlocal baryogenesis dominates, and if $G$-sphalerons do not violate
$L$, then one might face a variety of effects which suppress
baryogenesis~\cite{suppress}.

\subsection{Different transitions}

In many known intersecting braneworld models, the hidden sector
gauge groups are known to confine
(the $\beta$-function for the $USp$ groups are negative)~\cite{confine},
rather than break at low energies.
It is interesting to consider how this impacts baryogenesis.
The role of the first-order transition in the Sakharov conditions is
to drive the system away from thermal equilibrium.  From that point of
view, a first-order confining phase transition will do just as well as
a symmetry breaking transition.
The fundamental question is the suppression of
$G$-sphalerons after the transition.  In a Higgsing transition, it is
clear that $G$-sphalerons will be suppressed below the transition, and
thus would be unable to wash out the baryon asymmetry.  But after a
confining transition, it is not entirely clear if sphaleron-like
processes are suppressed.  This is analogous to the question of whether
or not strong sphalerons are suppressed at temperatures below
$\Lambda_{QCD}$.  It is of course difficult to make any
concrete calculations, due to the inherent difficulties in computing
in a strongly coupled gauge theory near confinement.  But we expect
that there should be a mass-gap on the confining side of the phase
transition.  Thus, we expect that there will be an upper limit on the
size of instantons after confinement.
As such, the energy barrier which the sphaleron-like
process must cross should have a non-zero minimum size, which in turn
implies Boltzmann suppression at low temperatures.  So although it is not
clear, it seems quite plausible that a first-order confining transition
in the $G$ sector can also produce a departure from equilibrium and seed
baryogenesis, while shutting off sphalerons to prevent washout.

\section{Signatures at the LHC}

It is of prime interest to determine the signatures for this type of
hidden sector baryogenesis (HSB) at LHC.  As mentioned, HSB
can occur even if $G$ breaks at a relatively high-scale, provided that
$B-L$ is also broken.  In this case, however, there will not necessarily
be any clear signature visible at LHC.  However, if supersymmetry breaking
is mediated to both the visible and hidden sectors by gravity, then one
might expect $G$ to in any case break at a scale $\sim \tev$.
One might expect that $G$-sphaleron processes can then be accessed at
LHC.  Unfortunately, this is likely not the case.  As shown
in~\cite{Dine:1989kt} in the context of electroweak theory, sphalerons
can be accessed efficiently at high temperature, but not in
high-energy scattering.  On the other hand, if $G$ sector particles
have masses set by the $\tev$ scale, then they can be produced
directly at LHC.

At the LHC, the exotic quark ($q$) can be pair-produced 
and the production process  in this case
would be $gg \rightarrow q{\bar q}$ via a $t$-channel exchange 
of the exotic quark. The exotic quark
$q$ would then decay into $q_{\rm  SM}$ and 
missing energy ($\eta$). The $q_{\rm SM}$ could be
one of the up type quarks. The Yukawa couplings between the 
exotic quarks and the SM quarks are
 controlled by the structure of intersections between the
$G$-branes and SM branes.
In general, the signal will be multiple jets +leptons(arising 
from the decay of top quarks) +missing energy.
  The exotic quark can also be singly produced via $gq\rightarrow q_{\rm exotic}
\eta$. The exotic quark then decays into a SM quark and $\eta$. The jet $E_T$ depends on
the mass difference between the exotic quark and $\eta$. If the $E_T$ is  large the signal
becomes more easily accessible. So the final
state can have a high $E_T$ jet plus missing energy.
We can also have leptonic signals once the $\lambda$s
are produced (via Z interaction) which will then decay into 
lepton plus missing energy ($\xi$). So the signal is
similar as in R-parity conserving SUSY
scenarios.

Interestingly, if the exotic quarks do not have the same hypercharge
as Standard Model quarks, then the scalars $\lambda$ and $\eta$ would
have fractional charge.  This would provide a unique signature of
new physics.  But due to the difficulty in decaying fractionally
charged particles into SM particles, such models would be tightly
bound by cosmology data and direct tests, requiring the fractionally
charged particles to recombine or annihilate almost entirely.

\section{Conclusions}

We have discussed a novel mechanism for baryogenesis which avoids
many of the tight constraints arising from electroweak and GUT baryogenesis.
This model utilizes a first-order transition in a hidden sector which
has a mixed anomaly with $U(1)_B$ to drive baryogenesis.  As such, this
mechanism naturally provides a way to break $B-L$, allowing the hidden
sector group to break at any scale without washout from electroweak
sphalerons (a major concern for GUT baryogenesis).  Furthermore, this
mechanism does not face the same challenge as electroweak baryogenesis in
fitting the precision data from LEP-II and other experiments.

Perhaps most notable, however, is that this mechanism seems natural in
IBM's.  Hidden sectors appear generically, and it is quite natural for
them to have mixed anomalies with $U(1)_B$.  For an IBM model to be viable,
such hidden sector groups (those with matter also charged under $SU(3)$)
must break, and if the breaking is a first-order transition then one would
expect baryogenesis.  This mechanism can be expected to occur quite naturally,
regardless of any other sources of baryon asymmetry.  If the exotic baryons
produced at the hidden sector transition can  decay to SM baryons, then HSB
can provide a substantial component of the asymmetry.  If not, then it will
provide exotic baryons which become a challenge in reconciling the IBM with
observation.
The signal of this scenario at the LHC will be consistent with multiple jets plus
missing energy and jets plus leptons plus
missing energy.

This is a fascinating example of how string theoretic input can provide
intuition for low-energy phenomenology and cosmology.  It will be interesting
to see how this mechanism works in specific models, particularly those IBMs
for which flux vacua can be counted.  It would be quite interesting to determine,
for example, brane models with large amounts of flux vacua which exhibit
HSB.  An analysis of the open string hidden-sector
landscape~\cite{Gomis:2005wc} would be quite useful for this purpose.

\section*{Acknowledgements}
We gratefully acknowledge Z. Chacko, M. Cvetic, J. Harvey,
D. Morrissey, A. Rajaraman, S. Thomas,
J. Wells and especially C. Wagner
for very helpful discussions.  J.K. would like to thank the Aspen
Center for Physics, where part of this work was completed, for their
hospitality.  This work is supported in part by
NSF grant PHY-0314712.

\end{document}